\documentclass[twocolumn,showpacs]{revtex4}
\usepackage{dcolumn,amssymb}
\usepackage{amsmath,epsfig}
\usepackage{rotating}

\newcommand{\bea}{\begin{eqnarray}}
\newcommand{\eea}{\end{eqnarray}}

\def\be#1\ee{\begin{equation}#1\end{equation}}

\def\be#1\ee{\begin{equation}#1\end{equation}}

\begin{document}

\title
{Simulating full QCD at nonzero density using the complex Langevin equation}

\author {D{\'e}nes Sexty}\email{d.sexty@thphys.uni-heidelberg.de}
\affiliation{Institut f\"ur Theoretische Physik, Universit\"at Heidelberg, 
Germany}
\date{\today}
\begin{abstract}
\noindent The complex Langevin method is extended to full QCD at non-zero
chemical potential. The 
use of gauge cooling stabilizes the simulations 
at small enough lattice spacings. At large fermion mass the results are 
compared to the HQCD approach, in which the 
spatial hoppings of fermionic variables are neglected, and 
good agreement is found.
The method allows simulations also at high densities, all the way up 
to saturation. 
\end{abstract}
\maketitle

The determination of the phase 
diagram of finite density QCD is one of the great problems of 
theoretical physics today. One is interested in averages defined with 
the Euclidean path integral
\bea 
\langle f[U] \rangle = {1 \over Z } \int DU e^{-S_g[U]} \det M(\mu,U) f[U], 
\eea
where $S_g[U]$ is the Yang-Mills action of the gauge fields and
$M(\mu,U)$ is the Dirac-matrix of the quark fields. 
Naive lattice simulations at $\mu\neq 0$ using importance 
sampling are made unfeasible
by the fact that the determinant of the fermion matrix is a complex number
in general. Various methods have been invented 
to circumvent the problem, but these are of limited use \cite{pdf}, mostly 
being applicable for $ \mu/T \lesssim 1 $. 
An exception is the complex Langevin method \cite{parisi}, which 
is not limited to small chemical potential.
 It has been demonstrated 
that this method allows for 
the solution of the sign problem in various systems \cite{aartsstam,Aarts:2008wh,Aarts:2010gr,aj,gaugecooling}, but in some cases 
also non-physical results are delivered \cite{Ambjorn:1985iw,Ambjorn:1986fz,aarts,Pawlowski:2013gag,Pawlowski:2013pje}.
 In this paper I demonstrate that 
the algorithm can be extended to full QCD with light 
quark masses on lattices with sufficiently small lattice spacings.

The complex Langevin method is based on setting up a complex Langevin 
equation (CLE) in an enlarged manifold, which is the complexification of 
the original field space \cite{parisi}. The original theory is 
recovered by taking expectation values of the analytically continued 
observables. 
For $SU(N)$ gauge theories 
this complexification
is $SL(N,\mathbb{C})$. This method can also be applied to other cases 
where the action becomes complex, e.g. the  case of real
time evolution, where the complexity of the action is much 'larger',
 using the Minkowskian formulation of the path integral 
\cite{bergesstam,bergessexty,opt}, or 
Yang-Mills theory with $\Theta$-term \cite{theta}.
 In this work I am concerned 
with finite density physics, where the complexity of the action 
is present at non-zero chemical potential.
The analytic understanding of the breakdowns and successes 
of the complex Langevin method has improved in the last few years \cite{ajss,guralnik,haarmeas,Duncan:2013wm,Aarts:2013uza}, 
one can gain an insight whether 
the results are trustworthy using requirements such as the fast decay of the 
distributions.

Recently an important breakthrough in this field was the development 
of a 'gauge cooling' algorithm for the CLE method \cite{gaugecooling}, where 
the gauge symmetry of the system is used to ensure a well localized
 distribution 
in the complexified field space, and thus convergence to the correct results. 

In this work the CLE method is applied to the lattice 
discretization of full QCD, i.e. for the action 
\bea S_{eff}[U] = S_g[U] - { N_F \over 4 } \textrm{ln}\, \textrm{det} M(\mu,U)
\label{qcdhatas}
\eea
where $S_g[U]$ is the Wilson plaquette action for the SU(3) link variables, 
and $M(\mu,U)$ is the unimproved staggered fermion determinant for 
$N_F$ fermion flavors 
\bea M(\mu,U)_{xy} &=& m \delta_{xy} + 
 \sum_\nu  { \eta_\nu(x) \over 2a}  \left[  { e^{ \delta_{\nu4}  \mu   }  
U_\nu(x) \delta_{x+a_\nu,y} } \right.  \nonumber \\  && 
- \left. e^{-\delta_{\nu4} \mu  }    
U^{-1}_\nu(x-a_\nu) \delta_{x-a_\nu,y} \right]    ,
\eea
where $x$ and $y$ indices represent spacetime coordinates,
 and $\eta_\mu(x)$ are the staggered sign functions. Periodic (antiperiodic) boundary conditions are used in space (time) directions.
 The fermion matrix fulfills the symmetry condition:
\bea
\epsilon_x M(\mu,U)_{xy} \epsilon_y = M^\dagger(-\mu^*,U)_{yx} 
\eea
with the ``staggered $\gamma_5$ matrix'', $\epsilon_x = (-1)^{x_1+x_2+x_3+x_4}$.
This symmetry leads to $ \det M(-\mu^*,U) = (\det M(\mu,U))^* $.
 This means that 
the determinant becomes complex for $ \textrm{Re}\, \mu\neq 0 $, making 
a simulation based on importance sampling impossible.  
Without rooting (i.e. using $N_F < 4$ by taking a root of the 
fermion determinant in the path integral), the staggered 
determinant describes 4 tastes
of fermions. In the Langevin dynamics (see below) the 
implementation of any (not necessarily integer) number 
of flavors is trivial, $N_F$ appears as a factor of a drift term. In this 
study I have chosen to use $N_F=4$ and $N_F=2$.

 The Langevin equation for the link variables is set up using the equation
\bea U_{x,\nu} (\tau+\epsilon) = 
  R_{x,\nu} (\tau) U_{x,\nu}(\tau),
\eea
with 
\bea
R_{x,\nu}(\tau) = \textrm{exp} \left[ i \sum\limits_a 
 \lambda_a ( \epsilon K_{ax\nu}  + \sqrt\epsilon \eta_{ax\nu} ) 
\right]
\eea
Here $\lambda_a$ are the generators of the gauge group, i.e. 
the Gell-Mann matrices. The drift force is determined by
\bea
K_{ax\nu}= -D_{ax\nu} S_{eff}[U]
\eea
with the left derivative
\bea
D_{ax\nu}f(U) = \left. \partial_\alpha f( e^{i \alpha \lambda_a} U_{x,\nu}) \right|_{\alpha=0}
\eea
The drift term for the action (\ref{qcdhatas}) is written as
\bea \label{driftterm}
 K_{ax\nu} &=& -D_{ax\nu} S_g [U]  \\ \nonumber 
 && + {N_F\over 4 }\textrm Tr[ M^{-1}(\mu,U) 
  D_{ax\nu} M (\mu,U)  ].
\eea
It has been suggested in \cite{logcut} that the drift corresponding 
to the action 
(\ref{qcdhatas}) might also include a term reflecting the branch cut of the 
complex logarithm on the negative real axis.
 This yet unclarified issue is the subject of 
ongoing research.

The drift term remains real only for $\mu=0$. Since the explicit calculation 
of the inverse of the fermion matrix is 
quite costly, this naive algorithm is feasible only for small lattice sizes.
As a cost effective 
alternative, the bilinear noise 
scheme \cite{batrouni,fukugita1987}, which is related 
to pseudofermionic variables, is introduced as follows.
The drift of the link variables is calculated using 
\bea
 K_{ax\nu} = -D_{ax\nu} S_g [U] + {N_F\over 4 } \eta^+ M^{-1} D_{ax\nu} M \eta,
\eea
where the $\eta$ is a vector of Gaussian random numbers 
satisfying $ \langle \eta^*_x \eta_y \rangle = \delta_{xy} $.
For the calculation of the drift term one has to solve the linear system 
of equation $ M^+ \psi = \eta$. In terms of the solution $\psi$ the 
drift term is written as 
\bea
 K_{ax\nu} = -D_{ax\nu} S_g [U] + {N_F\over 4 } \psi^+ D_{ax\nu} M \eta,
\eea
This means that this algorithm uses the conjugate gradient (CG) 
algorithm once for every timestep for the solution of 
the linear system. In \cite{fukugita1987} it was also shown that with a 
higher order algorithm
one can get rid of part of the $O(\epsilon^2)$ corrections from 
the Fokker-Planck equation.

From previous studies of the complex Langevin equation 
one learns the heuristic approach that a well localized distribution 
of the variables in the complexified field space is desirable.
A useful measure of the size of the distribution  in imaginary directions
of a link variable is the unitarity norm
\bea 
\textrm{Tr}(( U U^+ - 1 )^2) \ge 0,
\eea
where the equality is reached only for $SU(N)$ matrices.
The enlarged gauge symmetry of the system can be used to decrease the 
unitary norm of the system, thus ensure convergence to the exact results
\cite{opt}. Recently, we developed and tested 
in HQCD (see below) a procedure utilizing this freedom called gauge 
cooling\cite{gaugecooling} (reminiscent of stochastic gauge 
fixing\cite{stocgaugefix}). 
The idea is the following:
one uses gauge transformations 
\bea
U_{x,\nu} \rightarrow \Omega(x) U_{x,\nu} \Omega^{-1}(x+a_\nu)
\eea
with $ \Omega(x) \in SL(N,\mathbb{C}) $ to decrease the unitarity norm
of the system. This can be accomplished by choosing 
the $\Omega(x)$ matrices in the direction of the 
steepest descent of the unitarity norm.  Advanced versions 
of the algorithm ensuring faster decay of the unitarity norm use
 adaptive stepsize and Fourier acceleration 
\cite{Aarts:2013uxa}.

\begin{figure}
  \includegraphics[width= 0.9\columnwidth]{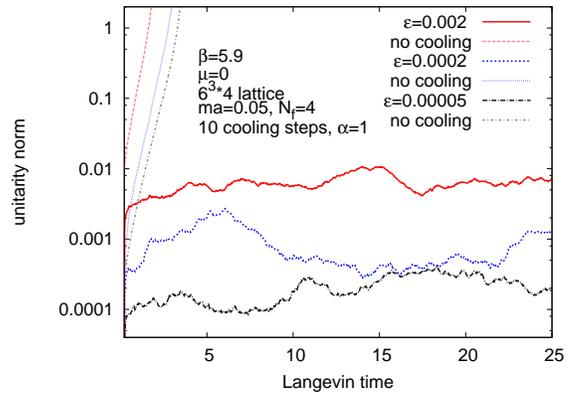}  
  \caption{ Unitarity norm as a function of Langevin time 
with and without cooling for several values of the Langevin 
timestep $\epsilon$.}
\label{cooledstep}
\end{figure}  

The consequence of using the bilinear noise scheme 
is that an imaginary part of the 
drift term is generated already at zero $\mu$, as the drift is  
real only on the average. Using a smaller Langevin step
allows the system to better approximate the drift term within 
a given Langevin time-window, therefore the resulting equilibrium 
unitarity norm of the simulation should vanish in the zero $\epsilon$ limit,
see Fig.~\ref{cooledstep}. Without gauge cooling the non-unitarities
generated by the noise term (or rounding 
errors in the case of the exact inverse algorithm) 
would grow exponentially, breaking down the simulation. 
Generally, we find also at $\mu \neq 0 $, using sufficient 
cooling, that the level of unitarity norm stabilizes and allows one to
obtain correct results (after $\epsilon \rightarrow 0 $ extrapolation)
for lattices with fine enough lattice spacings. 
As one observes, at low $\beta$ (below $\beta \approx 5.0-5.3$ for $N_F=4$) the 
cooling is not effective enough to prevent the system from wandering off far 
from the $SU(3)$ manifold, and 'skirted' distributions develop 
(as also observed in \cite{gaugecooling}). Close to the continuum 
limit, however, the algorithm seems to be stable irrespective of the physical phase, as observed using cheaper HQCD simulations.

\begin{figure}
  \includegraphics[width= 0.9\columnwidth]{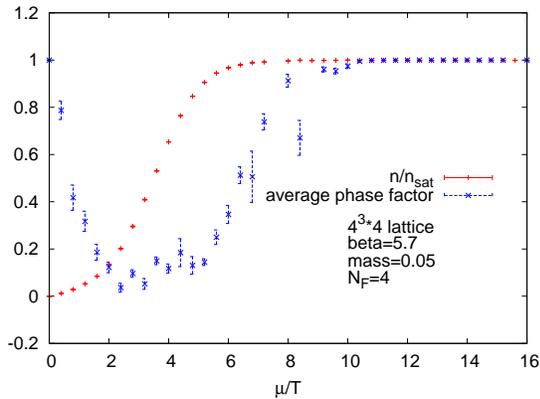}  
  \caption{ Average phase factor and density as a function 
of the chemical potential. }
\label{signdens}
\end{figure}

Observables are measured on 'slices' of the $ T,\mu$ phase 
diagram (meaning a scan using one variable while keeping the other fixed), 
to gain insight in the behavior of the system. On Fig.~\ref{signdens} 
a horizontal slice at high temperature is shown. The 
density of the fermions 
in the system is measured, as defined by
\bea
\langle n \rangle& =& {1\over \Omega} {\partial \ln Z \over \partial \mu },
\eea
 with $ \Omega$ the space-time volume,
in units of the saturation density (which is reached when all available 
fermionic states on the lattice are filled). 
The density starts to increase right away, there is no sign 
of the Silver-Blaze phenomenon \cite{silverblaze} at this high temperature, as expected.
 Around $\mu/T=8$ 
the saturation is reached.

To measure the importance of 
the fermionic contribution to the weight of the system,
we define the average sign of the determinant as 
\bea
\langle e^{2 i \varphi} \rangle = \left\langle {\det M( \mu) \over \det M(-\mu)} \right\rangle 
\eea
Since the calculation of the determinant is very costly, it is only 
measured on small lattices, see Fig.~\ref{signdens}. (For the Langevin 
dynamics the calculation of the determinant is not needed.)
One sees that even on this small lattice the average sign is close to 
zero in a big range of the physically interesting region, making 
reweighting unfeasible. (The feasibility of the reweighting 
is controlled by the sign average in the phasequenched system 
(where the determinant in the measure is substituted with its absolute value),
which behaves similarly to the sign average in the 
non-quenched system as shown in Fig.~\ref{signdens}.)
In the saturation region 
the phase fluctuations of the determinant vanish again, as the necessary 
energy to create a hole in the sea of fermions requires more
energy than is available in the thermal bath, thus the fermions 
decouple from the system.

\begin{figure}
  \includegraphics[width= 0.9\columnwidth]{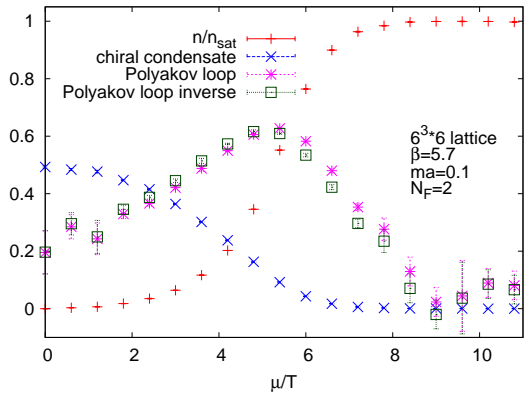}  
  \caption{ The fermion density, the chiral condensate (defined 
by $ \langle \partial \ln Z /\partial m \rangle / \Omega $)  and the 
trace of the Polyakov loop and its inverse as a function of 
the chemical potential. }
\label{horiz_slice}
\end{figure}

\begin{figure}[!ht]
  \includegraphics[width= 0.9\columnwidth]{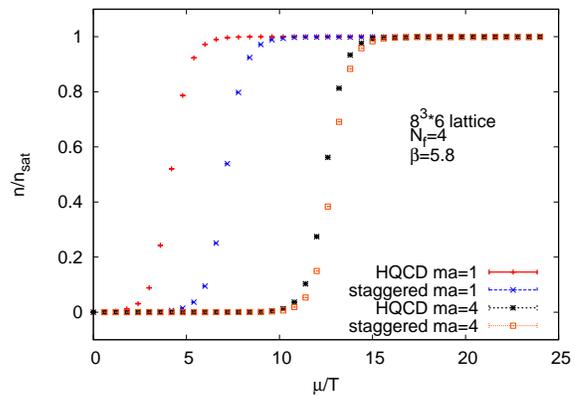}  
  \caption{ Comparison of the average densities 
 measured in HQCD and in full QCD with staggered fermions.}
\label{compplot1}
\end{figure}

In Fig.~\ref{horiz_slice} we show the fermionic observables 
as well as the trace of the Polyakov loops (defined in (\ref{poldef})) and its 
inverse again 
on a horizontal slice of the phase diagram.
The expected physical scenario is realized: the density of the 
fermions grows until saturation, while the chiral condensate vanishes.
The Polyakov loops have a peak at some nonzero $\mu$ 
(with the inverse Polyakov loop having a peak first), before they decay 
to zero, as the $Z_3$ symmetry of the system is restored in
the saturation region, where fermions no longer 
have an influence. Note that the critical $\beta$ of the system
(the value for which the system is at the transition between confined and 
deconfined phases)
is around $ \beta_c \approx 5.5 $ for the parameters
used in  Fig.~\ref{horiz_slice}, so the slice is slightly above  
the critical temperature. To reach smaller temperatures, lattices
using larger temporal extent are needed.

A well known approximation to full QCD is heavy 
quark QCD (HQCD), which is valid for heavy quarks and large chemical 
potentials \cite{oldhqcd,feo}, see also \cite{owe,deForcrand:2009dh}. In this approximation
the spatial hoppings are dropped and the 
fermionic determinant simplifies considerably:
\bea \label{hqdet}
\det (M(\mu,U)) = \prod\limits_x \det (1+C P_x ) \det (1 + C' P^{-1}_x)
\eea
with the Polyakov loop 
\bea \label{poldef}
 P_x = \prod\limits_{\tau=0}^{N_T-1} U_{(\tau,x),4}, 
\eea
and the parameters $ C= e^{\mu N_T} / (2 m)^{N_T} $ and 
$ C' = e^{-\mu N_T} / (2 m)^{N_T} $ with the staggered mass $m$, and 
the temporal extent of the lattice $N_T$. Note that 
this is the 'symmetrized' form of the determinant satisfying 
$ \det M(-\mu) = (\det M(\mu))^* $, otherwise the second factor could be 
dropped in the heavy-dense limit.
The corresponding approximation for Wilson fermions was 
studied with the 
complex Langevin method in an earlier 
publication \cite{gaugecooling}. The HQCD approximation for one flavor 
of Wilson fermion amounts 
to substituting $ m = 1/(4 \kappa)$ in eq.~(\ref{hqdet}),
as well as taking the square of the right hand side of (\ref{hqdet}).

\begin{figure}[!ht]
  \includegraphics[width= 0.9\columnwidth]{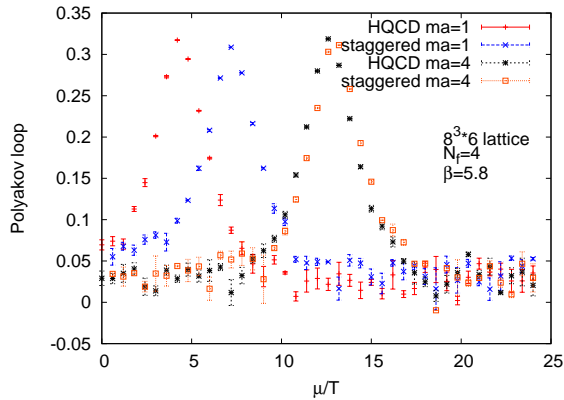}  
  \caption{ Comparison of the average
Polyakov loops 
 measured in HQCD and in full QCD with staggered fermions.}
\label{compplot2}
\end{figure}

Increasing the quark mass, the HQCD approach 
will become a 
better and better approximation of full QCD. To test 
at which mass scale will the HQCD 
become quantitatively accurate, and to validate the algorithm for 
full QCD, I compared simulations of HQCD
 (for details, see \cite{gaugecooling,Aarts:2013uxa})
to full QCD with staggered fermions, using the same mass parameter.
In Fig.~\ref{compplot1} the fermion density is compared,
in Fig.~\ref{compplot2}, the Polyakov loops are compared.

One observes good agreement at the high mass of $ a m=4$, as 
expected, since the HQCD expansion is based on the expansion of the 
fermion determinant using the small parameter $1/am$. 
 This of course does not prove that the 
results are fully reliable, but increases the confidence in the 
procedure, as the HQCD method was validated with reweighting at 
small $\mu$ \cite{gaugecooling}.
At smaller masses 
the results are quantitatively different, but the qualitative behavior
is very similar, where the biggest effect
on the density and on the Polyakov loop seems to be a rescaling 
of the chemical potential.

In this paper I have shown that finite density simulations of full 
QCD using the CLE with gauge cooling all the way up to
saturation are feasible using small enough lattice 
spacings, where the cooling is effective.
 This method avoids the sign and overlap
problems, direct simulation results in the high density region
are presented for the first time.
The cost of the simulation depends on the volume 
similarly to a hybrid Monte Carlo simulation, as the inversion of the fermion
matrix is the main numerical cost. In particular the cost  
increases polynomially with the volume, in contrast with 
the exponentially costly reweighting approach.

The results correctly reproduce the saturation physics 
and are found to agree with HQCD for large quark masses. 
To increase the confidence in the reliability of the 
results, further checks are needed in the regions where different 
approaches are available, such as 
results at small chemical potentials \cite{compare}, or the results
gained using strong coupling expansions.


{\it Acknowledgments.} -- I am indebted to Gert Aarts, 
Erhard Seiler and Ion-Olimpiu Stamatescu for many discussions 
and collaboration on related work. 
A large part of the numerical calculations for this project was done on the 
bwGRiD (http://www.bw-grid.de), member of the German D-Grid initiative, 
funded by BMBF and MWFK Baden-W\"urttemberg.

\end{document}